# "Against Marrying a Stranger:"
# Marital Matchmaking Technologies in Saudi Arabia

**Adel Al-Dawood**
University of Minnesota
Minneapolis, MN
aldaw004@umn.edu

**Norah Abokhodair**
University of Washington
Seattle, WA
noraha@uw.edu

**Houda El Mimouni**
Drexel University
Philadelphia, PA
he52@drexel.edu

**Svetlana Yarosh**
University of Minnesota
Minneapolis, MN
lana@umn.edu

## ABSTRACT
Websites and applications that match and connect individuals for romantic purposes are commonly used in the Western world. However, there have not been many previous investigations focusing on cultural factors that affect the adoption of similar technologies in religiously conservative non-Western cultures. In this study, we examine the socio-technical and cultural factors that influence the perceptions and use of matchmaking technologies in Saudi Arabia. We report the methods and findings of interviews with 18 Saudi nationals (nine males and nine females) with diverse demographics and backgrounds. We provide qualitatively generated insights into the major themes reported by our participants related to the common approaches to matchmaking, the current role of technology, and concerns regarding matchmaking technologies in this cultural context. We relate these themes to specific implications for designing marital matchmaking technologies in Saudi Arabia and we outline opportunities for future investigations.

## ACM Classification Keywords
H.5.m. Information interfaces and presentation (e.g., HCI): Miscellaneous; K.4 Computers and Society.

## MOTIVATIONAL SCENARIO AND INTRODUCTION
Abdullah is a 29-year-old Saudi male. He approaches his mom to ask her to engage him to his childhood crush, but his mom is not as motivated as he is. Abdullah's crush becomes engaged to someone else, leaving him frustrated and heartbroken. Through the connections of his mom and sisters, he tries multiple times to find another suitable spouse, but is frustrated because his family often decides based on their own perceptions and without consulting him. Even in cases where a potential match is suggested, the prospective bride's family makes it hard for him to get any information about her, unless they meet in chaperoned settings. Abdullah has no alternative but to go through his mom or sisters, which is the most socially accepted method to finding a spouse in Saudi Arabia. Abdullah wants to have agency in this process and believes that technology may support his search. However, he knows that his culture considers "online dating" inappropriate, which leaves him stuck!

Online personals have become prevalent for finding a spouse [15] and dating in many parts of the world [32]. Yet, these technologies have largely been rejected by Saudi Arabian culture. Take for instance, the number of Saudis subscribed to Muslima.com, one of the biggest Muslim matrimonial websites, is one of the lowest in Muslim countries [24]. This is due to many factors: First, it reflects the traditional and religious conservative culture of Saudi Arabia, especially with their concerns for privacy [3]. Second, there is social stigma associated with participating in non-traditional behaviors (e.g., dating) [26]. Third, there is a common belief that Saudis who use these sites do so to engage in culturally controversial acts (e.g., flirting, attending mixed-gender gatherings, etc.) Thus, seeking a marriage match using non-traditional routes may affect one's reputation, which is a high-stake value in this context, even if one's intentions are sincere. Furthermore, the Human-Computer Interaction (HCI) community has not investigated this context, as most matchmaking technologies have focused on the concerns, values, and priorities of cultures where dating, flirting and mixed-gender gatherings are the norm and accepted.

We address this gap by addressing the following questions:

RQ1. What challenges do Saudis face with the traditional matchmaking approaches?

RQ2. What are the perceptions of technologically mediated matchmaking in Saudi Arabian culture?

RQ3. What are the opportunities for culturally-appropriate technical interventions in this context?

We investigated these questions through the analysis of 18 in-depth interviews with Saudi participants with the aim to contribute to the ongoing discussion in HCI on inclusive design and cross-cultural computing.

**STUDY CONTEXT: CHARACTERISTICS OF SAUDI ARABIA'S CULTURE VIS-À-VIS MATCHMAKING**

To understand the context presented in this paper, it is important to explain some of the dominant cultural aspects of Saudi Arabia that are related to marriage and matchmaking. We highlight the following aspects: influence of gender-segregation, the current marriage process, family values, and technology use vis-à-vis marriage.

**Gender Segregation in Saudi Arabia**

The segregation of opposite genders in Saudi Arabia is an important aspect. Males and females who are not *mahrams*– unmarriageable kin (e.g., mothers, sisters, and aunts)–are not permitted to mix or be together alone [4] for non-professional purposes. Schools, hospitals, banks and other public spaces are designed with gender segregation in mind [3, 4]. This general rule influences what is considered appropriate and dictates the public and the private sphere. Moreover, wedding ceremonies in Saudi Arabia are segregated, where the males and females will have their own independent ceremonies. Occasionally, unrelated males and females can mix for professional purposes, such as, conducting a business or seeing a doctor of the opposite gender. However, even within these interactions certain norms are enacted and no physical contact takes place. As a result, gender-segregation is one of the primary societal aspects that influences the conduct of marriage, as males and females have limited opportunities to meet before they are officially engaged [7]. Dating, thus, is also forbidden.

**Stages of Traditional Matchmaking**

To date, in Saudi Arabia, traditional matchmaking, or arranged marriage, is considered the most common way for finding a life partner [29, 37]. Accordingly, males rely on the guidance of their female *mahram* to find and recommend suitable females for marriage. Females, on the other hand, are on the receiving side, as they await to be recommended or seen by other females who are looking for potential brides. Wedding parties and social gatherings are considered a great place to be seen and sought after and for other females to find a bride for their male relatives.

In an arranged marriage, there are a few stages that need to be fulfilled before the marriage occurs. In the following we highlight the most important and common stages:

1- The Proposal: A senior female leader of the prospective groom contacts the mother or senior female member of the prospective female with the intention of proposing. After the initial agreement between the two families takes place, the prospective groom and bride enter the *Social engagement* ("*kh'otobah,*") stage. This stage is like "dating" in a Western sense. During this time the couple gets to know each other by talking on the phone and through chaperoned face-to-face meetings. Still, the fiancé is not yet considered a *mahram* and the fiancée must maintain the Islamic headscarf or "*hijab*" during their meetings.

2- The Contract: if the couple are happy with each other and are ready for the next step, they announce what is called the *Religious engagement* ("*a'ked qaran* or *melkah*.) During this stage the couple will have a *Shaikh* (Muslim cleric) to marry them by signing the marriage contract after the marriage conditions have been met, including the payment of a dowry *(mahr)* in the form of money or valuable goods (such as gold jewelry) to the bride declaring the marriage. After that, the couple is considered legally married, the bride can remove her hijab and meetings can be unchaperoned. However, in most cases, they still do not live together until the wedding ceremony.

3- The Wedding ceremony: a celebration (*'irs or zaffaf*) is the last stage, and is usually celebrated with a wedding ceremony. The goal of the ceremony is to serve as a public announcement of the marriage.

**Family Dynamics and Values**

In Islam, "*silat ur-rahm*" is an expression that means maintaining family relationships. Muslims are required to sustain a good relationship with their parents, siblings, and extended relatives. This relationship is based on love, respect, and care. Obeying and being kind to one's parents is an Islamic obligation. These are considered guiding values within this context and are mentioned in many versus in the Quran. The importance of these values have generated a society that is highly collectivist in nature [7, 33]. This makes it important that spouses get along with their in-laws to preserve a good relationship between the families. These values are very prevalent in the region and dictate familial relationships. Therefore, the opinion of family members and relatives on who one marries is respected and considered.

**Permissibility of Using Technology for Marriage**

For Muslim clerics, the permissibility of using technology for ordinary life matters is still a topic of debate due to the ongoing challenge of defining the line between proper an improper use. The general guideline is that if the use of technology is beneficial to Islam and to oneself, it is permissible. Otherwise, it is considered to be a waste of time and may lead to immorality [42]. Understandably, there is room for divergent interpretations, which allows some flexibility and assumes a person is accountable for their actions [42]. Specifically with cyber dating, the general guidelines, as mentioned by Wheeler [42], are that:

1) You have the intention to marry and communicate to know each other better;
2) It is done in a respectful and moral way;
3) Your parents are informed;
4) You do not delay marriage more than needed.

. It is worth noting that the older generation who still controls the marriage process prefer traditional methods versus modern ones (i.e., using technology.) This is slowly changing with the recent trend of using social media to evaluate potential spouses in Saudi.

Little previous work has investigated *the de facto* use of matchmaking technologies in Saudi Arabian culture—a gap we address with this work.

## RELATED WORK

We describe relevant work in two areas: technology-mediated matchmaking and cross-cultural research.

### Technology Mediated Matchmaking

Researchers have explored many areas of online dating and technology-mediated matchmaking (e.g., self-presentation [12, 17, 18, 21, 23, 34, 39, 40], mate preferences [14, 19, 25, 35], deception [13, 18, 22, 39, 40]).

Ellison, Heino and Gibbs [12] found that participants mediated the tension between impression management and the desire to present an authentic sense of self. This was through tactics such as creating a profile that reflected their "ideal self." Hancock and Toma [23] examined the accuracy of 54 online dating photographs posted by heterosexual daters. While online daters rated their photos as relatively accurate, independent judges rated approximately 1/3 of the photographs as not accurate. In this study, we approach a unique cultural context, where interaction between genders is not widely accepted and females are not expected to share their real photos on social media [6, 9, 20].

Toma and Hancock [39] highlighted the association between attractiveness and deception. They also linked such association to the technological affordances that allow online daters to engage in selective self-presentation. In addition, Rosen, et al. [34] compared between online and traditional dating. They found that the amount of emotionality and self-disclosure affected a person's perception of a potential partner. An e-mail with strong emotional words (e.g., excited, wonderful) led to more positive impressions than an e-mail with fewer strong emotional words (e.g., happy, fine). This resulted in nearly three out of four subjects selecting the e-mailer with strong emotional words for the fictitious dater of the opposite sex.

Lastly, Fiore and Donath [14], revealed interesting results that illustrate how the medium shapes the way daters present themselves. The study found that users of an online dating system looked for people similar to themselves. The information online dating systems provided differed greatly from what a person might gain from a face-to-face interaction.

Our work provokes the question of what changes in the current methods used for matrimonial matchmaking are needed for matchmaking technology to be more accepted in these contexts. This question is becoming more relevant as social media becomes more prevalent in Saudi culture.

### Cross-Cultural Research on Mediated Relationships

HCI literature has chiefly focused on how technology is used in cultures where premarital relationships are the norm. Technology designs informed by these assumptions reflect these cultural values. Comparatively few studies have focused specifically on how technology is used in the Arab world and how this different set of values may impact design.

Elmasry [31] analyzed how different people used Facebook and Hatfield and Rapson [24] examined how culture shapes the use of technology in the Middle East. Guta [20] assesses how Saudi females use social media to negotiate and express their identity, which included finding a spouse. In the context of dating, Alsheikh, Rode and Lindley [7] investigated how 11 Arab individuals use technology in the context of their long distance romantic relationships. Lastly, Abokhodair [3] explains how privacy is perceived differently in the Arab Gulf and how it plays a role in the use of technology. The findings highlighted key differences that match with cultural expectations that impacted the use of technology.

Saudi's conservative culture has mostly followed traditional matchmaking through personal connections [11]. With social media becoming more prevalent [24] and improved awareness of potential health issues of marrying relatives, marrying non-relatives has increased [41]. Nevertheless, issues such as trust [5], privacy [1–3], and maintaining long distance relationships [7] manifest as roadblocks to accepting the use of technology for matchmaking.

The above studies [7, 24, 31] give a general idea how culture shapes technology use. Elmasry [31] compared the users from liberal to conservative (US being liberal, Qatar conservative and Egypt somewhere in between). Other studies paint all the Arab/Islamic countries with a wide brush [7, 24]. They do not provide a contrast of the diverse subcultures in the Arab world. Further, none of the studies focused on conservative populations in the Arab/Islamic culture using technology for matchmaking. We address this gap in our work.

## METHODS

### Recruitment

The first author, who is Saudi, initially recruited study participants through Saudi friends and an online sign up form that was shared through many social media platforms. The downside of this method was concerning reaching enough female participants with diverse backgrounds. This was solved by asking a female Saudi public figure on Twitter to retweet the online sign up form. We aimed to recruit participants in equal proportions from the three major regions of Saudi that are represented in the following cities: Riyadh (the capital), Jeddah, and Dammam. In addition, three age groups were represented for each gender: 18-25, 25-35 and 35-50. Most of the participants were between 25 and 35 years old (N=10), and participants from the age range of 35-50 were a minority in our study (N=3). We also recruited participants from all marital statuses, including single, married, and divorced, achieving a balanced distribution of each.

We continued recruiting participants until we deemed that data saturation was reached (i.e., most themes were repeating in the data). While we estimated that saturation was reached after 14-16 participants, we continued with a few additional interviews to confirm that no new themes were emerging and ensure enough diversity within participants' demographics. That been said, we acknowledge that the recruitment process is subject to self-selection bias and thus our findings might not be representative of the whole population of Saudi Arabia.

**Interview Procedure**
We conducted in-depth interviews with 18 participants, 9 males and 9 females. Table 1 provides additional details about each participant. Interviews were conducted mostly in-person or through Skype, but some were also done over the phone. Interviews were conducted in Arabic (a native language for most of the authors) and took between 60-90 minutes on average. Participants were offered compensation for their time, in the form of a 15$ gift card or cash either in Saudi riyals or US dollars. But many participants were happy to take part in the study voluntarily and rejected the monetary compensation. Our semi-structured protocol included questions on the following major topics of interest: participants background, thoughts about current marriage methods, how can technology play a role, reflections on previous experiences if any and conclude with general remarks about the subjects. Additionally, participants provided written consent and basic demographics using an online form.

| Participant Codes | M/F | Age Bracket | Marriage Status |
|---|---|---|---|
| p1 | M | 35-50 | Married |
| p2 | M | 25-35 | Single |
| p3, p4, p12 | M | 18-25 | Single |
| p5, p6, p7 | M | 25-35 | Married |
| p8 | M | 35-50 | Remarried |
| p9 | F | 18-25 | Engaged |
| p10 | F | 18-25 | Single |
| p11 | F | 25-35 | Divorced |
| p13, p14, p16 | F | 25-35 | Single |
| p15 | F | 35-50 | Divorced |
| p17, p18 | F | 25-35 | Married |

**Table 1. Study participant characteristics.**

**Transcribing, Open Coding, and Affinity Diagraming**
All interviews were transcribed in Arabic by the first author who then open coded the transcripts (following the protocol described in [27]). The generated codes were then carefully translated to English to support the collaboration between the authors during the axial coding stage, memo writing, and clustering process. During the transcribing and open coding stages, we took into consideration the context by explicating the meaning in relationship to the context rather than a mere focus on the literal. This was especially the case when idioms were used or terms that have a certain meaning in Saudi Arabian dialogue. After open coding, affinity diagraming was used to enact constant comparison between open codes (as described in [30]) to generate more thematic clusters. In this work, we focus on themes most relevant to the perceptions and use of matchmaking technologies in Saudi Arabia.

**Ethical Considerations in a Male Guardianship Culture**
There were unique ethical considerations in this work that may be informative to others doing research in this cultural context. As Saudi Arabia is a male guardianship culture, the Institutional Review Board(IRB) requested that we take special care with recruiting and interviewing female participants. Our initial recruitment was limited to males to allow us to refine the protocol and identify any potential cultural concerns. Before proceeding to interviewing females, the IRB's full board reviewed the protocol and initial insights at the lead author's institution. The final approved recruitment call requested that female participants obtain permission from their male guardian to take part in the study (though it did not require us to solicit any evidence of this approval). We explained to each female participant that should their male guardian object to their participation, we would be obligated to remove their data from the record. The inclusion of this request led to unfortunate side effects, as the *de facto* reality of Saudi Arabian gender politics do not always align with the *de jure* laws that bind IRB's policies. Our call caused several Saudi women to contact the study PI directly with concerns that requiring male guardian permission endorsed problematic gender practices. It is worth noting, that during the time the call was posted on Twitter, an ongoing social media campaign requiring the Saudi government to lift the male guardianship law was simultaneously occurring [38]. In response, we added a footnote to the call, explaining the specific reasons for the consent procedure and disavowing any endorsement of specific male guardianship laws. Other researchers may experience similar challenges if attempting to run studies in a similar context. Addressing these challenges is imperative, as it is important to engage the voices of women in formative research and design process.

**RESULTS**
We present our themes in three major clusters: current issues with the marriage process, opportunities for technology, and technology fears and concerns.

**Current Issues**
In discussing the topic of marriage and finding potential partners with our participants, we learned that there are issues that might stem from family or society expectations. We divide these issues into two categories. The first category concerns the issues around finding potential partners and the societal and religious barriers. Particularly, the family's role in marriage and gender segregation, which limits interactions between potential partners. The second category concerns the practical and institutional obstacles that can limit one's options and choices for marriage. Examples include costs and expenses of marriage, linage concerns, and intercultural marriages.

*Family Role*
Participants discussed the importance of the family in the process of finding a potential partner as explained in the following excerpt by P4. He is studying in the US and hopes to break out of the norms and marry a non-Saudi female as he hinted having relationships with females in the US. He says:

> *Maybe [parents] think they understand life and have lived it and thus must enforce it on [their children]. That they see the best and suitable for them. For example, a mother would say to her son 'this [wife] is suitable for you' or ask her son 'you want to marry? [The family] will look for you'.*

The common belief is that mothers (or any close female relative like a sister or an aunt) will know better than anyone else who the right partner for their child is. Because she is the one who knows them well and has the knowledge and experience of being married. However, in some cases the interference of the family and the pressure they put on their children can lead to issues. P18, who met her husband through her blog and then got in touch on Facebook, explains the issue in the following excerpt, she says:

> *Methods used now that cause problems or embarrassment is concerned with the intolerable interference from the family. For example, while looking for a life partner and when we are trying to know them there is a lot of pressure from family. I think this sometimes causes negative consequences because the person cannot be relaxed while under a lot of pressure.*

In this example, P18 talks about the lack of freedom and flexibility in getting to know her potential partner during the *family engagement* phase. She mentions that the family pressure that follows can lead to "negative consequences," which can range from the breaking of the engagement to divorce in some extreme cases. This is normally due to couples not given enough time to get to know each other in different aspects of life. She elaborates on this point by saying that *"during engagement the male and female try to hide all their flaws and show their best and when marriage comes they show their true colors and collision occurs."* Meaning that the husband and wife will face reality once they are married and live together. It is worth noting that during the family engagement couples do not meet alone and the bride's father or brother (male *mahram*) chaperon meetings. However, they can speak on the phone, and that is what P18 is referring to in the previous excerpt. She emphasizes that on the phone you cannot know much about your partner and you get to face reality when one is already married.

Overall, the interviews participants' opinions varied on a spectrum from ones who support and encourage family involvement and traditional marriages to others that wanted to find their own partners outside the control of the family. P11, who felt culture and customs limited her ability to know her ex-husband enough, comments on this practice: *"Males and females need to have awareness that nobody has a say in their choice [of whom to marry.]"* In this excerpt P11, is clear about her message regarding the need to increase awareness for both genders to have the freedom of choosing their own partner. On the other hand, P7, who married the daughter of his dad's friend, suggests a middle ground approach. He says: *"taking your own opinion is important but also see your family's opinion, especially, your mom and dad."* This relates to the fact that marriages usually unite two families together. Spouses not getting along with their in-laws is not usually a desired outcome by both the spouses and their families.

*Gender Segregation and Desegregation*
In Saudi Arabia, gender segregation is culturally enforced; it is not accepted for non-*mahram* males and females to mingle together in public places. Because of gender segregation, it becomes very hard for males and females to meet or find potential partners. Things are changing in the Kingdom, however. A push towards modernization and globalization is slowly transforming the traditional social norms. Big international companies and hospitals are now moving towards a controlled version of a gender-mixed environment. But, these workplaces are still limited to males and females who come from families that do not mind mixed work environments. Commenting on this issue, P5, who studied in the U.S before and after marriage, says:

> *When I came to America barriers were broken. It was like I was in Saudi but in a mix gendered world. It benefitted me more because I got to interact with a lot of Saudi females.*

Opposite to what he has experienced in Saudi Arabia, in the U.S. P5 got to meet and interact with many Saudi females studying abroad at his university and found that to be a positive experience. On the other hand, P15, felt that social pressure did not allow her to talk enough with her ex-husband. She comments on the drawbacks of gender segregation as an issue that effects females more than males, she says:

> *Our society has not reach the point where a female can meet, go out and get to know someone…[we lack] freedom for a female to have her own opinion.*

In this excerpt, P15 shares her frustration regarding the issue of social norms preventing her from getting to know her future husband at the time of the interview, before being married to him. On the other hand, P9 was expressing how some Saudis found a way to work around to meet potential partners and make it look like it was done traditionally:

> *"Society doesn't have to know that I met a guy online…a girl went to a wedding that the guy's family attended to make it as if they coincidently met." P9*

She later expresses her disapproval of it *"I don't like these ways and feel they are twisted a bit."*

*Costs and Expenses*
Participants expressed a concern with the expensive marriage costs in Saudi Arabia. In general, wedding ceremonies in the Arab world are known to be associated with prestige and recognition, particularly on the bride's side. The tradition is to invite relatives, extended family, and friends, so an average wedding will host around 200 guests. Hospitality is shown through the amount of food offered and the type of entertainment the hosts of the wedding provide. Different regions in Saudi have different traditions and expectations about who hosts and organizes the wedding (whether the bride or groom.) On average, marriage costs start at 150,000 riyals (about $40,000), including the *mahar,* the wedding ceremony, gifts, and housing [36]. The highly-exaggerated cost of getting married was a topic of concern to many of our participants. P9 comments about this issue:

> *[Exaggerated wedding costs] are not supposed to be accepted. I ask myself that these elders, maybe those responsible for the*

*female or legislators in the society, how do they accept these [exaggerated wedding costs] and not reject them.*

Another participant, P7, recommends government intervention on this issue. He says: *"government should intervene to place limits on mahar [and wedding costs] so it does not become and obstacle and place punishments if they do not follow them."* P8, a remarried male participant, expresses his aversion to the lavish customs, he says:

> There are some customs I hate, such as, when you get married, you should rent out a Lexus (an expensive car brand), a hotel and get flashy. The honeymoon must be a big deal.

*Lineage and Intercultural Marriages*

Saudi Arabia encourages within-Saudi marriages and makes marriage hard for males and females who want to marry from outside the Kingdom or the Gulf Cooperation Council (GCC). There are certain regulations and lengthy procedures that Saudis marrying from outside must adhere to before getting married, including a government approval that might take up to a year with stringent conditions to get the approval. It is worth noting that the approval still does not grant the citizenship to the foreigner spouse nor to the children of a Saudi female marrying a foreigner. Thus, families in many cases are discouraged from approving inter-country marriages for their children, especially their daughters. Expressing his opinion on this issue, P1, which has been married traditionally for almost 10 years, says:

> I see it harder because you will most likely end up with a person who is not close to the family or tribal society you live in and know their customs and culture. You should be careful because you do not want your relationship to end in divorce.

Saudi Arabia is considered a tribal society, therefore, maintaining blood ties and tribal traditions are of high importance to many families. Many cases have been reported of the challenges that surround males and females marrying from outside their tribes. There seems to be a voice within society that believes it is an issue that needs to get resolved as expressed by P10 *"Ashraaf (people with direct blood-line descendant from Prophet Muhammad) only marry Ashraf and Bedouins only marry Bedouins. This is a wrong concept."* P9 can relate more to this issue as her fiancé is of a tribal linage while she is not. Even though her family was understanding and accepted her fiancé, for the couple, life ahead is filled with struggles:

> We feel it is hard for our families to understand each other…His family accepted me and my family accepted him but to deal with it later is not guaranteed…two different societies.

**Opportunities for Technology**

While not all issues that are present in the marriage process are related to technology, participants have provided a lot of insight on the ways technology can play a role in the process. The role can be either by complementing the current process, offering ways to keep in touch, or helping to get to know more about their potential partners. Also, technology can fill in the gaps for those who cannot utilize traditional methods since they depend completely on personal connections. Lastly, developing technology that is culturally sensitive will be accepted and welcomed by the society that will allow it to be a vehicle for change.

*Enhancing Traditional Matchmaking*

Given that traditional marriage is by far the most reliable method due to the importance of the *"trust factor through family,"* as stated by P2. Yet, P2 believes that *"social media complements,"* and can make this process better. In addition, P8, who went through three family-arranged marriages before finding his current wife, suggests that technology can broaden the search horizon since traditional means are limited to personal connections. P8 argues: *"Your family might search [for a bride from] 50 families during 10 years…through the database you have a million families to choose from."* P8 believes that the traditional method is inefficient as it only allows him to meet a limited number of potential wives and is time intensive. He believes technology will allow him to reach more potential partners in a short period.

As mentioned earlier, due to additional challenge of gender segregation, knowing one's future partner becomes harder. However, social media is increasingly offering young males and females a new means for connecting and bending societal constrains. This is embodied in the following statement by P5: *"If the female had an account you can see what she cooked or if she made a painting… things that [will] keep you informed [about her]."* P5 believes that the ability to view a female's public profile in social media will allow him to know more about her. Since the traditional method limits the amount of information about her, P5 finds social media the perfect medium to fill in the gaps and know more about her interests and hobbies.

Earlier research has discussed the common issue of data accuracy on social media. Due to the nature of these platforms, it becomes easy for the user to provide an inaccurate persona [10]. This could occur in the form of users manipulating their profile to present themselves in a specific way[10]. However, this still does not devalue the experience. P6 states that: *"She might delete or clean up [her profile] …[but it] at least gives you some hint about what her interests are."* P14 explains how she interacts and presents herself on twitter in the following passage:

> View his tweets and respond to them…show my interests…usually it is showoff…I put more literately quotes…it might have principles that could represent you in a good way.

Our participants mentioned that social media allows them a real-time natural observation of others online participation such as comments, likes, or posts. However, this is limited by the profile being public as stated by P7: *"I might have a public account that has things I care about, such as art and books."*

Since the interaction online is somewhat indirect, it allows for easy exploration of others without worrying about formalities that are required in the traditional method. P18 explains how her husband learned more about her from her Facebook profile: *"He went to my Facebook page…so he can decide if I am suitable or not without him needing to ask a*

*person."* Her husband found her through her blog and enjoyed reading her blog posts. That motivated him to check her Facebook profile to know more about her before talking to his family to propose to her.

Certain technologies like Match.com seem to be promising, but do not adapt to cultural differences very well as they were designed with a specific group of users in mind (i.e. Western populations). One of the important culturally sensitive design needs for a website like Match.com to fit in the Saudi context is regulating the process. That will allow it to be trusted and decrease harassment. Harassment in the context of Saudi Arabia can include the case when someone pursues the other with no serious intention to marry, but rather for dating or being friends. This is vaguely mention by P18 as *"things they do not want"* in the following excerpt:

> *A platform that is suitable for our customs…like Match.com, there would be moderators…it would be highly regulated so people can trust that when they enter this place they are not exposed to harassment or things they don't want.*

Certain populations might benefit from technology because of their limited or non-existent social connections, which traditional methods are based on. For example, P5 talks about an acquaintance struggling to get married because *"he does not have sisters"*, which means he does not have anyone to look for him. P1 brings up another example and that is for orphan females who do not have male relatives or family to represent them appropriately *"there are many deprived females because there is not anybody who can reach them in an appropriate way."* Another rare case mentioned by P5 is when the female does not go to many social events with her mother *"sometimes the mom does not take her daughter to weddings and funerals."* This is because social events in Saudi are the main space for females to be seen and known to other families that may consider her as a bride for their son or refer her as a bride for distant relatives or friends.

### Knowing Each Other

A major technology that seems to be used by many in Saudi Arabia is Twitter, producing about 150 million tweets a month [8]. They use it to interact with others, which may not be possible in real life. This is mainly important because it is viewed as a space to exchange and interacte with others' ideas and opinions. P9, who met her fiancé through Twitter, states: *"Twitter allows the most interaction with people as ideas."* She believes Twitter allowed her to understand others intellectually. Others prefer Facebook, as they find it more inclusive. P17, who got to meet and know her husband through Facebook, says: *"it allows for many things to see in a person."* She believes that Facebook is not limited to a certain medium (e.g. text, image, and video) but rather allows expressing in multiple ways.

Regardless of the technology being used, P15 believes that technology in general allows for a flexible interaction with the opposite gender with no strings attached: *"Technology allows talking comfortably and [if things are not working] separation is easier."* A more detailed description on how technology can mediate the interaction that would be otherwise unacceptable because of cultural barriers is explained by P5:

> *Applications that show you points on a map that is big enough to show individuals that you can tell that this person represents it…assuming society accepts it…he sends to her that I want to propose to you. She sees the person, she does not like him and sends to him to get lost.*

P5 believes that direct interaction might not be acceptable, but an indirect interaction with physical presence might be a middle ground that could be accepted. Even when interaction through technology is considered acceptable, having it regulated by a mediator may make it more meaningful for those involved as P8 describes *"a three person skype, me…a mediator and the female…[mediator] runs the discussion."*

Many participants believed going through conflicts or problems would help understand each other's thought process. In addition to seeing their potential partner in different circumstances, which might be closer to reality. These could arise over time as events occur naturally per P14 *"in a year conflicts arise…conflicts show the other face"* or caused intentionally by one of the partners as P3 insists *"I have to create problems in this period…see the reaction of the other in front of me."* P6 suggests a different method for testing your partner, this is to involve a third party to arrange it: *"put them in a dilemma…and they decide if they are suitable."* Although this is not a safe playground, there is always the issue of balance—that is, being realistic and reaching a decision in a reasonable period.

While physical appearance may be an important factor when considering a significant other, P9 disagrees and believes it is important to know the mentality of your future partner: *"When you feel harmonized intellectually, looks are not important."* P18 considers physical appearances to be misleading and can cause intellectual compatibility to be overlooked *"Create attraction… [That] does not have intellectual compatibility."* She believes it makes her focus on physical attraction without noticing intellectual incompatibly. P16 agrees that intellectual attraction should be the main attraction *"Be more intellectual and not have emotional be a big part."* This might align well with some cultural expectations of females not showing a personal picture of themselves as P18 states *"If people accept that she puts her picture."*

### Quality of Interaction

Most participants agreed that nothing could compare to meeting your potential partner in person. P3 mentions that *"seeing your facial expressions while talking"* is very important for him. However, when meeting in person is not feasible, technology might be suitable to simulate the interaction per P7 *"I take pictures of my daily life…It is like you are virtually with the person."* P12 offers a solution that was outside the box for us, that is the use of holograms to simulate a virtual reality scenario for both partners as if they are next to each other to simulate the feeling of being together. There was the concern of it being acceptable as it may be considered a physical mixing of genders that is not allowed religiously:

> *Holograms…there would be a scenario…You see how she moves, how she looks, her actions…like virtual reality…I do not know if it would be mixing or not.*

Another aspect is the issue of transparency in regards to the traditional method, P6 explains: *"the problem of the traditional method is the lying and deceit."* This is caused by the mediator, which is usually the mother as P9 states: *"the information provided by the mothers about her children is filled with lies,"* which is why P9 prefers online means to traditional ones. P18 believes it may come from the desire of both parties to advance beyond the engagement period: *"during engagement the male and female try to hide all their flaws and show their best."* P17 disagrees because it might end up as an emotional relationship too quickly *"it should not transform quickly into an emotional relationship."* P8 thinks indirect interaction might allow more transparency *"A person is not shy to say something in text like they are shy to say it on the phone."* P7 suggests that anonymity could be helpful *"Ask relatives or friends in an anonymous way,"* since judgment is usually reduced relatively.

**Technology Concerns and Fears**

While technology represents an opportunity for match seekers, certain fears and concerns are overtly expressed. Specifically, trusting the unknown, the seriousness of online match seekers and customs violation.

*Trust when Using Technology*

In the absence of strategies and devices that enable match-seekers to distinguish lies from facts, establishing trust remains a challenge. P9 expressed worries about approaching someone for marriage without knowing the person previously:

> *Lie detecting technology…it is better to have a measure to confirm that they are truthful with you because you have not seen them and do not know who they are.*

The issue of trust is common in the literature. However, in the case of Saudi Arabian culture, trust does not stem only from the fact that the two parts have not met in person before but also from the fact that establishing a relation needs to come through common connections (i.e., referrals). P14 stressed the point that a successful and a sustainable marriage requires that either the two families know each other or the mediator is known by both families:

> *Who led him? There has to be previous knowledge or…people that know you…If there is not it would be a problem for the female. There will be collision, there will be complete rejection.*

Participants expressed worry regarding the fact that marriage seekers can hide the dark side of their lives to show a deceiving image of a perfect life. To this point, P15 posits: *"see the background of the person…it is important for me to know things that are disastrous."*

Nevertheless, as match-seekers require sharing negative experiences, the issue of privacy emerges and thus the need for privacy management becomes crucial. P2 wonders: *"I do not know if there is a mechanism to share your info without violating your privacy."* In the same regard, P2 wanted to manage his profile in a way to share only with the person he is interested in, but still maintain his privacy. Also, he believes it is important to ensure that both are only considering one potential partner at a time rather that considering many at the same time:

> *If there is a way to ensure a one-way communication…I could communicate with everyone but with the person I want to know more, I want to focus on him or her.*

*Serious Use of Technology*

Some participants questioned the usefulness and usability of technology for seeking a life partner. For them marriage is a very serious matter that cannot be tackled through a website or an app. P13 believes social media is not an appropriate tool for seeking marriage. It does not align with Saudi culture and how marriage in such context is carried out. She believes technology should be utilized appropriately:

> *Social media is not suitable because it does not match the way of thinking in our community…Technology is very beneficial but we do not use it for this issue*

In the same token, P15 questions the seriousness of those people using technology for marriage and relates to her experience with a male that was not serious and could have led to him exploiting her *"he was not serious and gave room that it becomes exploitive".* Also, P17, who met her husband online, mentions that the process is complicated and is not necessarily easier than the traditional method; *"It is not easy to find someone to marry through the internet. It is really hard and a very complicated process."*

*Technology Might Violate Customs or Religion*

Because Saudis believe in gender segregation and protecting females from stranger males, online match seeking is considered an outlying behavior that contradicts the common prevailing cultural and/or religious rules. P18 illustrates such common viewpoint:

> *The internet made it easier for youth to know each other but I am against knowing each other when they are complete strangers.*

In the same vein, P13 explains the reason behind such position and refers to the fact that in online match making there is no initial agreement. This is comparing to what happens in the real life Saudi Arabian marriage process. Such agreement indicates the seriousness of the marriage seeker and protects the females. The above condition is believed to be a religious one which stands highly above any customs or cultural considerations:

> *If we allow them to know each other before there is initial agreement, a lot will not allow it because we always talk and say it is prohibited in religion…we are not limited by customs and culture but religion and we have mostly 90% conservatives*

An interesting point relevant to the Saudi Arabian context is the nuances in what interactions between the potential partners can be allowed during the matchmaking process. P14 highlights such element:

> *Some families allow phone calls before legal marriage, but in our community, they still reject this. Like there is no communication until the day of the meeting.*

With that being said, the permissibility of exchanging phone calls, images and conversations using technology prior to any formal meeting in the Saudi Arabian society remains relative. Nevertheless, it is rare for families to allow interaction between partners before they are engaged in the Saudi Arabian culture.

## DISCUSSION

### Unique Considerations for Technology Use in Conservative Societies

This investigation reveals the unique aspects of the marriage-seeking process in Saudi Arabian culture. It may align with conservative traditions in other Muslim countries, but it differs from the Western process in significant ways. Saudi Arabia's conservative culture has mostly followed traditional matchmaking through personal connections [11]. As online dating became popular, technology has provided a novel opportunity for marriage seekers. However, most available marriage platforms do not account for the unique context of Saudi Arabian marriage, thus the numbers of Saudi's using such platforms remains low.

The following cultural elements constitute the unique aspects of Saudi Arabian match making culture and represent roadblocks to Saudis' use of the available technology for match seeking purposes. 1. *Gender segregation:* a woman and a stranger man are not allowed to meet without a *"mahram"* for the woman to be protected. In contrast, online dating platforms are not designed for gender segregation or the involvement of *mahram* chaperones. 2. *Family involvement: "bir al walidayn"* is a rooted principle stipulating that a son or a daughter need to involve their parents in the marriage process and obey their opinions. Online dating platforms currently do not afford this kind of parental involvement. 3. *Khotobah*: The formal agreement between both the families that is necessary for the relationship to proceed. This notion is absent in online dating, thus exchanging phone calls and messages violates customs or religion in conservative cultures like Saudi Arabia.

The absence of such cultural elements in the design of match making platforms involves three values of ethical importance [16]: trust, privacy and credibility. (1) While trust as a value has been frequently highlighted in works relevant to matchmaking and dating [21, 40], trust in this Saudi Arabian context is violated by the fact that no family member knows the other party or a third party mediating the introductions. This is exacerbated by the fact that people tend to misrepresent personal characteristics in online dating contexts [23]. (2) Privacy is also a common issue specifically in the Saudi Arabian context where women are encouraged to be covered from head to toe. Striking a balance between trust and privacy is a difficult goal since gaining trust presupposes revealing more about the self, which has the potential for endangering privacy. (3) Lastly, our study reveals that the credibility of technology for matchmaking is questioned by Saudis. The fact that existing platforms do not comply with the Saudi culture stigmatizes the entire class of technologies, associating it with culturally forbidden activities like flirting or fornication.

### Opportunities for Future Investigation

Many concerns for transparency and honesty have been expressed by participants, which may align with the work done by Ma *et al.* [28]. It is worth investigating how transparency and honesty can be attained while still preserving Saudi Arabian values. Our participants mentioned many existing technologies, like Match.com, Twitter, and Facebook. As we seek a cross-cultural understanding of social computing, it would be prudent to conduct in-depth studies of how these technologies are used, their pros and cons, and whether they can be appropriated in culturally sensitive ways (or if new technologies need to be made available in these cultural contexts). We discuss these tensions below.

*Appropriating Current Platforms or Developing New Ones*

Our participants have expressed interest in certain technologies (e.g., Facebook, Twitter, Match.com) to find potential partners. However, it is still unclear whether the best solution is to design and develop a new platform or support appropriation of present platforms. This is a question we also aim to tackle in our future work. Specific concerns with current matchmaking technologies could be addressed by implementing affordances for gender segregation and family involvement, which are important in Saudi Arabian culture. Our future research will focus on developing and deploying culturally-sensitive prototypes to understand how these implications may be embodied in technical artifacts. In addition, we will investigate the role of matching algorithms, to understand how profile information and personalized matching may be designed to accommodate Saudi Arabian cultural values and Muslim users' in general. Namely, we plan to interview Saudis about using technologies like Match.com and how they can be designed to be culturally appropriate.

*Designing for Traditional versus Emerging Values*

Many HCI researchers have offered ways to keep users' values in mind during the design process. One of the well-known methods is Value Sensitive Design (VSD), which is a "theoretically grounded approach to the design of technology that accounts for human values in a principled and comprehensive manner throughout the design process"[16]. One of the decisions that VSD keeps open to the designer is regarding the nature of the values they are designing for and whether the designer role is to remain neutral or partisan in the selection of values. In the case of Saudi Arabia, the research to date is revealing a set of complex values that lean towards traditional values, rather than emergent ones.

We learned from our participants that the most applicable design for this context will allow them to benefit from modern technologies while maintaining traditions, customs, and religious practices that they value. However, these preferences may not be representative of all Saudi Arabians and may privilege certain value systems. Therefore, as we continue in our research, we leave a set of open questions for the DIS community, that is, in terms of values and design, what values do we design for? And how do we avoid marginalizing a group of users with design? How do we design systems that can respect the needs of cultural traditions without creating systems that reinforce existing power structures? As we struggle with our stance as designers in this context, we invite the larger DIS community to participate in this conversation.

**LIMITATIONS**

The type of methods used and participants we were able to interview limited the study. While we recruited for diverse participants that represented different subcultures within Saudi, we neither sought nor attained a representative sample. Several self-selection biases may have influenced our recruitment. For example, a male conducting the interviews limited female participants to those who self-selected as being comfortable talking to a male about the topic. We also noticed that most participants seemed to have progressive views, which may not be representative of the older population. Finally, the interview method relies on self-report by the participant. Participants (especially, female participants) may have felt pressure to self-present in a positive light and focus on views consistent with their cultural context. We hope that in the future these limitations may be ameliorated through investigations of this topic using other methods and triangulating findings between this and other work.

**CONCLUSION**

In this paper, we highlighted the unique context of Saudi Arabian marital matchmaking. Through an analysis of in-depth interviews with 18 Saudis, we uncovered the culturally accepted process of marital matchmaking, discussed the potential role of technology, and highlighted concerns in this context. We found that while participants wanted to benefit from matchmaking technologies, they were only willing to use such technologies if they fitted within their cultural context and adhered to their religious norms. The findings of this study support the growing interest within the HCI community in culturally-sensitive and inclusive design that focuses on understanding and accounting for human values in the design process to allow for proper deployment and global use. We also provided design principles to support culturally-sensitive design of matchmaking technologies in this context, focusing on providing affordances for gender segregation and family involvement. Finally, we uncovered several challenges for researchers and designers working in this context and provided our solution. With this work, we provided an additional contextually-grounded research study to benefit the future value-sensitive work in the DIS community.

**ACKNOWLEDGMENTS**

We would like to thank all participants for their time and sharing their private lives with us. Special thanks for Eman Al-Dawood for her early translations of consent forms during IRB procedures. Also, Loujain Alhathloul helped greatly with recruitment of Saudi females and bringing attention to their concerns. Lastly, a token of appreciation goes to Jennifer Rode for connecting the 1st author with collaborators.